# Towards an Open Science Platform for the Evaluation of Data Fusion

Weinan Huang[1], Junyi Chen[1], Lei Meng[1], and David Lillis[1,2]

[1] Beijing Dublin International College
University College Dublin
Dublin, Ireland
{weinan.huang,junyi.chen,lei.meng}@ucdconnect.ie

[2] School of Computer Science
University College Dublin
Dublin, Ireland
david.lillis@ucd.ie

*Abstract*— **Combining the results of different search engines in order to improve upon their performance has been the subject of many research papers. This has become known as the "Data Fusion" task, and has great promise in dealing with the vast quantity of unstructured textual data that is a feature of many Big Data scenarios. However, no universally-accepted evaluation methodology has emerged in the community. This makes it difficult to make meaningful comparisons between the various proposed techniques from reading the literature alone. Variations in the datasets, metrics, and baseline results have all contributed to this difficulty.**

**This paper argues that a more unified approach is required, and that a centralised software platform should be developed to aid researchers in making comparisons between their algorithms and others. The desirable qualities of such a system have been identified and proposed, and an early prototype has been developed. Re-implementing algorithms published by other researchers is a great burden on those proposing new techniques. The prototype system has the potential to greatly reduce this burden and thus encourage more comparable results being generated and published more easily.**

*Keywords*- information retrieval, data fusion, evaluation

## I. INTRODUCTION

Dealing with vast quantities of unstructured, primarily textual data is a key issue in Big Data research. May different architectures and algorithms have been proposed to tackle the Information Retrieval (IR) task: identifying information items that satisfy a user's information need. One approach is to use "data fusion", where the results of multiple individual IR systems are merged to form a single set of results to present to the user [1]. Data fusion specifically refers to the situation where these individual systems search the same corpus.

Although numerous researchers have addressed data fusion, little consensus has emerged as to which algorithms and approaches perform best in a given situation. This paper has two primary aims. Firstly it seeks to establish a need for standardised evaluation practice in this area of research by reviewing and critiquing evaluation approaches used in the literature (Section II).

This discussion provides the motivation for second primary aim: the proposal of a standard experimental platform for conducting data fusion experiments. Section III outlines the desirable features of such a platform. An initial prototype platform incorporating the fundamental features has been developed. Its current state and future aims are presented in Sections IV and V respectively. Finally, the paper concludes in Section VI.

## II. BACKGROUND AND MOTIVATION

A "data fusion" system is intended to merge the outputs of individual "component systems", each of which produces a ranked list of documents (or other information items) in response to some query. This process is based in the idea that a combination of results from effective individual search engines will intuitively result in a higher-quality result set for presentation to the user. This is considered to be preferable to displaying several distinct ranked lists, which will likely overlap to some degree, to a user.

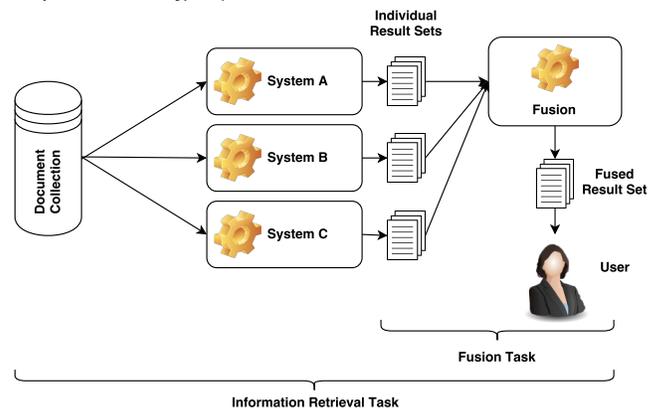

Figure 1. The fusion process.

The process required for the data fusion task is illustrated in Fig. 1. After receiving a query from a user, this shows that a set of heterogeneous component systems retrieve documents from the same document collection in response to this query. Each system produces a result set that consists of a ranked list of documents that it considers to be relevant to the query. These are then merged using some fusion technique, to produce a single set of results for display to the user.

From the user's perspective, this process is no different to the traditional IR task, where a query is provided and a single set of results is returned. The fusion task, however, is only part of this process, taking the outputs of existing systems as its inputs, in order to create the final result.

Due to the IR community's long-standing tradition of creating and distributing standard document collections, queries and relevance judgments, there is little need for fusion researchers to create their own document collections and/or component systems. The Text REtrieval Conference (TREC)[1] has long been a rich source of experimental data for IR researchers. For the purposes of data fusion, TREC is even more important, as it additionally publishes the result sets submitted by various research groups in tackling particular tasks. Where these tasks are ad hoc in nature, these result sets can then be used for fusion. This advantage is reflected in the popularity of TREC data in a large quantity of data fusion experiments, as evidenced in [2-15]. In addition, each TREC "run" includes responses to multiple standard queries (typically 50) from a single system. Thus the fusion of multiple systems' outputs can be conducted using their responses to the same queries.

ImageCLEF[2] uses a similar approach to TREC, albeit the task is limited to image retrieval. Data from this conference has also been used for fusion [16].

Despite the availability of datasets that could reasonably be described as being "standard", the literature shows little consistency in the choice of specific inputs to the fusion process. There has historically been a tendency towards the use of the most recent suitable data, which makes comparison to older papers difficult. Also, each track offers numerous runs provided by different participating groups, of which any combination may be chosen for fusion.

Aside from the different choices taken with regard to input data, the following sections also describe how they have generally been deployed in an inconsistent, incomparable way across the data fusion community. Section II-A demonstrates how the community has not settled on standard metrics to be used in the evaluation of fusion tasks. Following this Sections II-B and II-C show how the choice of baseline comparison is similarly nonstandard. Other inconsistencies surround the use (and non-use) of statistical significance tests (Section II-D) and training data (Section II-E).

### A. Use of standard IR metrics

A feature of most data fusion research is the use of standard IR metrics for evaluation. This is in keeping with other areas of IR research. Popular metrics include:
- Precision at *n* documents (P@n) [3,5-7,9,11,12,14,15,17,18].
- Average Precision (AP) or Mean Average Precision (MAP) [2-18].
- Precision/Recall curves [4,10,12].
- Binary preference (bpref) [6,7,12,18].
- Recall-level precision (R-prec) [13,15].

Generally, the choice of metrics reflects the wider IR community. The popularity of P@n and MAP is in keeping with wider trends. However, as the following sections discuss, even the choosing the same metric does not necessarily make experiment comparable. Also, changing trends in IR make comparisons with older papers difficult. For example, NDCG has not been popularly used for fusion, likely due to its relative newness. Although it was initially proposed in 2002 [19], its popularity within the IR community emerged later.

### B. Comparison with component systems

Many researchers compare the fusion output with the quality of its inputs as part of their evaluation. This is intuitively reasonable, as it acknowledges that merging result sets does not guarantee an improvement in quality. In the majority of cases, the comparison is made with the component system that has achieved the highest score according to the metric being used [2,4,5,10,13–17]. A fusion method that improves upon this justifies its development by indicating that it is not possible to achieve equivalent results simply by choosing a single high-performing IR system. Fusion is only useful if it can improve upon the available component systems. Alternative comparisons include those involving the mean evaluation scores of the component systems [2,14] or the median run amongst those available [9].

However, this comparison with component systems is not uniformly applied, as evidenced by published studies that lack this type of evaluation [3,6,11,12].

### C. Comparison with other fusion algorithms

In addition to comparing a proposed fusion algorithm with the component systems' outputs, evaluation typically also includes a comparison with competing data fusion algorithms. This can be seen in [2,4–15,15–17,20].

While there is great variation in the algorithms chosen for comparison, two common baselines are the CombMNZ and CombSum fusion algorithms [21]. These are attractive as they have been shown to perform well despite being conceptually simple and easily implemented [20]. Typically, newer techniques are shown to perform favourably compared to these algorithms, and this is claimed as a validation of their worth. However, it is far less common to see comparisons made with these later approaches themselves. This is possibly due to the fact that more sophisticated techniques are less easily implemented or that open source implementations of more complex techniques tend not to be available. Even when two novel techniques perform favourably compared to CombMNZ or CombSum using a standard metric such as MAP, the differences between the datasets used render direct comparisons impossible without further information.

### D. Statistical significance tests

As most single-score evaluation metrics are averaged over multiple queries, many researchers also include statistical significance tests. Without this, a few outlier queries may cause the mean scores to increase even if the more common scenario is for performance to disimprove, or vice versa. For this purpose, a two-tailed paired t-test is most commonly employed [3,5–8,11–16]. The Wilcoxon signed-rank test has also been used [9,17], although it has been stated that this test is not suitable for IR [22].

---
[1] http://trec.nist.gov
[2] http://imageclef.org

*E. Training data*

Some data fusion approaches rely on knowledge of the past performance of the component systems. Where this is the case, evaluations generally follow the lead of the machine learning community by dividing the set of available queries into a training set and a test set [2,6–8,12,16,23]. The proportion that is used for training varies across experiments, with no particular approach appearing dominant.

One observation to be made here is that even in situations where the same runs from the same task are used for evaluation in different papers, an experiment that uses a subset of the available queries for training will not be comparable with one that uses all available queries.

## III. AIMS OF AN EXPERIMENTAL PLATFORM

As discussed in Section II, the comparability of data fusion approaches is compromised by the lack of consistency in experimental design. To combat this problem, this paper proposes an experimental platform to harmonise the evaluation of data fusion approaches. An early prototype has been developed, which includes many of the fundamental features that are necessary for such a platform. This type of approach would not be feasible for the general ad hoc IR task, since the datasets are very large and there is huge variability in the choice of libraries, tools and programming languages that are employed to tackle ad hoc tasks. In contrast, fusion requires standard inputs that are comparatively short (the longest TREC runs consist of 50,000 lines: a maximum of 10,000 results for each of 50 queries). Also, the algorithms themselves are less complex, and are consequently more amenable to being ported to a common programming language, without requiring external libraries that may be platform-specific.

This section outlines the long-terms goals of the project, as well as discussing the challenges that need to be overcome in order to achieve each of these goals. Each goal is rooted in a desire to enable comparable experiments to be run, by standardising the methodology, data and algorithm implementation being used.

**Centralisation:** An evaluation platform should be centralised, and be made available through the web. This way, all researchers have access to design new experiments, as well as to examine the results of past experiments.

**Datasets:** Of fundamental importance is the availability of standard datasets. This applies not only to the choice of appropriate TREC tracks, but also the decision as to which runs and queries to use for experimentation. Where the data is from third party sources (e.g. TREC), it is essential to be mindful of copyright issues.

**Implementations:** Standard implementations of baseline algorithms must be available also. In most cases, algorithms are described only mathematically, and it has not been common for source code to be made available in a complementary way. The platform would allow authors to claim ownership over the implementation of an algorithm, and provide a definitive implementation if desired. This helps to address situations where some factors (e.g. optimal values for certain parameters) are not explicitly mentioned in published papers.

**Storage:** Historical experiments must be maintained for future reference. This avoids a situation where individual researchers are required to re-run past experiments in order to make comparisons with new approaches. Instead, an implementation of a new approach can be provided, which is then asked to fuse results based on existing experiment setups. The results of this can then easily be compared with results from techniques that are already on record.

**Evaluations:** It is important that a standard evaluation practice be established, based on community-adopted norms, such as the trec_eval evaluation tool. As noted in Section II-A, particular metrics fall in and out of favour over time, and therefore this aim is intrinsically linked with the idea of storing results of past experiments. Where a new evaluation metric is proposed, storage of historical experimental results makes it a trivial task to re-evaluate outputs of various techniques using a new metric. This also has the effect of future-proofing experimental results.

**Visualisations:** As the primary aim of proposed platform is for it to be used by researchers in the area of data fusion, it is important that appropriate visualisations be made available to aid with the interpretation of results.

**Export:** Users will wish to publish the results of their experiments, and so the details of these must be exportable in a format appropriate for publication. This can be in the form of graphs, but also as tables formatted using LaTeX, so that they can be directly incorporated into papers. This also has the effect of promoting consistency across the presentation of results for future experiments.

**Consistency:** In any experimental setup, it is crucial that only comparable results are used to draw conclusions. Where one algorithm requires a subset of the available queries for training data, it must be the case that these queries are ignored by any technique that does not rely on training data. This thus ensures that the test queries are consistent for all techniques being compared. Cross validation of the type used as standard in the machine learning community is useful for fusion also.

## IV. PROTOTYPE EXPERIMENTAL PLATFORM

An initial prototype has been developed to provide the most fundamental features of those desired. This prototype is a web-based system developed in Python using the Django framework[3]. Python's pandas library[4] was used for data manipulation and plotting graphs for visualisation.

Implementations of a number of existing fusion algorithms were written in Python. These followed a standard pattern so that additional implementations of other techniques can be seamlessly added at a later date. To date, the algorithms implemented include simple interleaving, CombMNZ [21], CombSum [21], Linear Combination (with manual weighting) [23], ProbFuse [6], SegFuse [12] and SlideFuse [7].

---

[3] http://djangoproject.com
[4] http://pandas.pydata.org

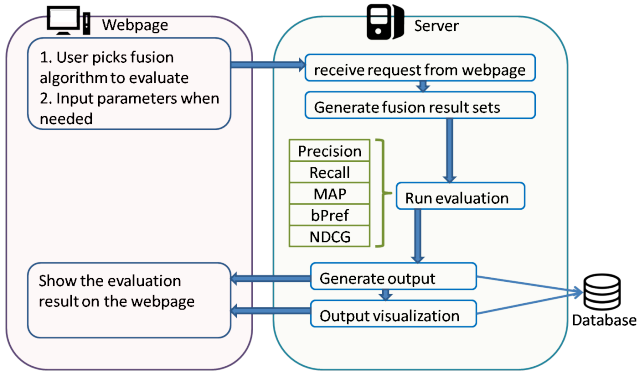

Figure 2. Flow chart for prototype platform.

Implementations of standard evaluation metrics were written in Python also. These were tested by comparison with the output of trec_eval. The metrics implemented were precision, recall, MAP, bpref and NDCG.

One common method of comparing the outputs of these metrics is the use of percentage increases and decreases [2,3,7,9,13]. It was decided not to include this method of comparison on the basis that it can be heavily biased by the level of the baseline being compared.

Experiments are conducted and stored automatically. Users must choose the algorithm(s), dataset(s) and evaluation metric(s) to use. This is done by presenting the users with buttons representing those that are available. Next, the system will prompt the user for any parameters that are required, after which the experiment can be run. Parameters include those specific to particular algorithms (e.g. SlideFuse requires a specific window size, and a linear combination requires weights for each input system to be provided), but also those that relate to the experiment at a whole (e.g. which queries should be used as training data for algorithms that require it).

Fig. 2 shows a flow chart illustrating how the prototype functions. Both the fused results and the evaluation results are stored in a MariaDB database[5]. Visualisation of results is provided upon conclusion of the experiment for easy comparison. The evaluation results are presented in both tabular and graph format. The graphs are generated using the Matplotlib library.

## V. LIMITATIONS AND FUTURE WORK

As discussed in Section II, the ultimate motivation of this platform is to support researchers to publish comparative studies in academic venues. To achieve this goal, further features will be required, so as to achieve all the aims outlined in Section III. This section outlines the future work that is planned for this project so that the platform can be adopted in the field of data fusion.

Firstly, user registration and permissions are important. It cannot be the case that any user can simply upload an implementation of a fusion algorithm. This will likely result in several competing and incompatible implementations of the same algorithms being uploaded to the system. The original authors of an algorithm should be supported in providing an "official" implementation of their approach. Also, although the existing implementations have been created in a standard way, it is not yet supported that new implementations can be uploaded via the web interface.

---

[5] An open-source SQL-based database system: http://mariadb.org

The second major feature that is necessary is the ability to add a new algorithm to an existing stored experiment. This is a key vision of the experimental platform, to ease the evaluation of new approaches, and to provide uniformity between the experiments that are chosen for publication.

The dataset used in the prototype is currently limited to the TREC 2014 Web Track results. To date this is the latest year in which a large-scale ad hoc retrieval task was included as part of TREC. This will be extended further to include a variety of result sets.

Further, more minor features, that are planned include:
- Export of comparison tables in LaTeX format for direct use in publications.
- Export of fused result sets in TREC format, so as to be usable with trec_eval and other tools.
- Export of zipped archives of graphs.
- Statistical significance testing for evaluation results.
- Development of alternative, fusion-specific evaluation metrics to address the fact that obtaining perfect scores on traditional ad hoc metrics is frequently impossible.

## VI. CONCLUSIONS

As can be seen from the comprehensive literature review presented in Section II, the methods of evaluation used for data fusion do not lend themselves to easy comparability between experiments. Thus, despite many years of work in this area, it is difficult to reach an informed opinion as to the relative merits of competing techniques. This is because of a wide variation in the evaluation methodologies, metrics, datasets, and baselines that have been employed in the literature.

This paper argues that data fusion is an appropriate forum for the implementation of a standardised experimentation and evaluation framework. Because the nature is of the task is relatively standardised, being the merging of ranked lists of results, it requires much less variation than standard IR or data analytics problems. Section III sets out the desirable attributes of such a system, with a view to outlining what will be necessary to achieve this goal. An early prototype that provides the key features has been developed, and has been outlined in Section IV. This does not yet satisfy all the desirable properties, and further work is planned so as to make a tool that provides value, verifiability and comparability to the Information Retrieval community. Providing a centralised service to the community has the potentially enhance the overall comparability of studies published in the data fusion domain, leading to a better overall understanding of the relative performance characteristics of proposed algorithms. In the longer term, this will help developers of Big Data systems involving unstructured textual data to make a more informed decision about the choice of data fusion algorithms that are available.